\documentclass{UoBnote}
\usepackage[usenames,dvipsnames]{color}     
\usepackage{verbatim}
\usepackage{url}
\usepackage[latin1]{inputenc}
\usepackage[T1]{fontenc}
\usepackage{amsmath, amssymb}
\usepackage{subfigure}
\usepackage{xspace}
\usepackage{float}
\usepackage[page,toc]{appendix}
\usepackage{multirow}
\newcommand{\Finesse}{\textsc{Finesse}\xspace}
\usepackage[numbib]{tocbibind}

\usepackage{fancyvrb}
\DefineVerbatimEnvironment{finesse}{Verbatim}
{formatcom=\small}

\author{Stefan Ballmer, Jerome Degallaix, Andreas Freise, Paul Fulda}
\shorttitle{\Finesse vs. OSCAR vs. Analytic Alignment}
\title{Comparing \textsc{Finesse} simulations, analytical solutions
  and OSCAR simulations of Fabry-Perot alignment signals}
\date{\today}
\issue{2}
\ligodcc{T1300345}
\usepackage[pdftex,pagebackref=true,pdfpagelabels=true]{hyperref}
\definecolor{linkcolor}{rgb}{.8,0,0}
\definecolor{urlcolor}{rgb}{0,0,.7}
\definecolor{citecolor}{rgb}{0,.5,0}
\definecolor{acrocolor}{rgb}{0,0,.7}
\hypersetup{bookmarksopen,colorlinks=true}
\hypersetup{pdfstartview=FitH}
\hypersetup{linktocpage=true,bookmarksnumbered=true}
\hypersetup{plainpages=false,breaklinks=true}
\hypersetup{linkcolor=linkcolor,citecolor=citecolor,urlcolor=urlcolor}

\begin{document}
\maketitle
\tableofcontents
\vspace{1cm}\hrule \vspace{1cm}
\clearpage

\section{Introduction}
This document records the results of a comparison the interferometer
simulation \Finesse~\cite{finesse} against an analytic (MATLAB based) calculation of
the alignment sensing signals of a Fabry Perot cavity. This task
was started during the commissioning workshop at the LIGO
Livingston site between the 28.1.2013 and 1.2.2013~\cite{workshop} with the aim
of creating a reference example for validating numerical simulation
tools. The FFT based simulation OSCAR~\cite{oscar} joined the battle later.

\section{The test setup}

\begin{figure}[H]
\centering
\includegraphics{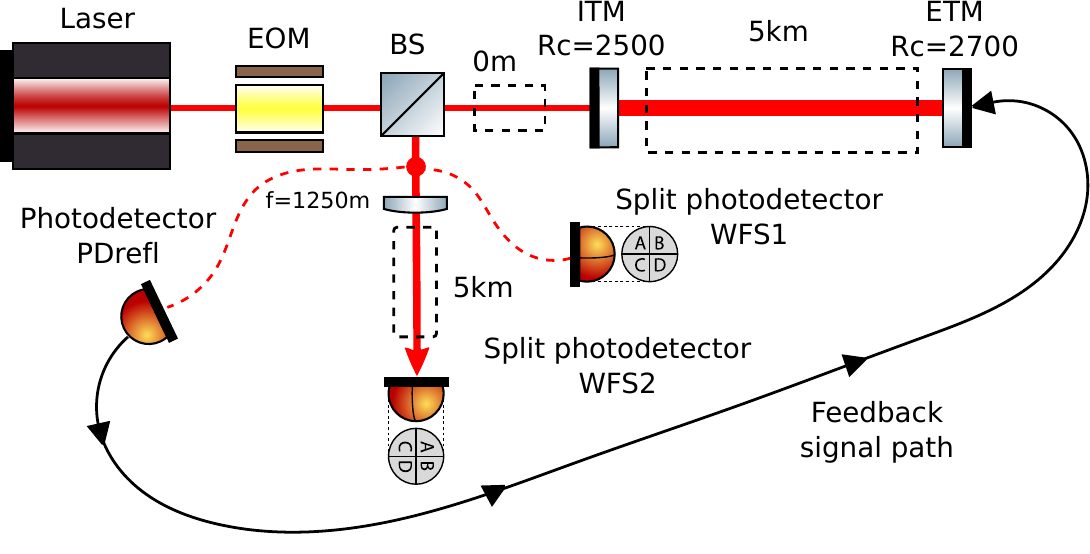}
\caption{Schematic of the test setup modelled throughout the document:
a linear optical cavity is used with input laser, EOM and several
photo detectors to investigate the wavefront sensing signals in the
case of thermal distortion of the input mirror ITM.}\label{fig:setup}
\end{figure}

The basic setup is a linear Fabry Perot cavity and a phase modulated
input beam. The reflected light is detected by two wavefront sensors
(WFSs). WFS1 is located directly at the input mirror's front surface,
a second beam is directed via a pick-off mirror, a lens and
a 5\,km distance towards WFS2. The setup is shown in
figure~\ref{fig:setup} and the main parameters are given
in table~\ref{tab:params}.

\begin{table}[bh]
\centering
\begin{tabular}{l}
\hline
laser power $P=1$\,W\\
laser wavelength $\lambda=1064$\,nm \\
EOM frequency  $f=9$\, MHz \\
EOM modulation depth $m=0.001$ (as in $\phi(t)=m\cos(\omega_m t)$) \\
ITM $Rc=2500\,{\mathrm m}, R= 98\%$, $T=2\%$\\
ETM $Rc=2700\,{\mathrm m}, R= 99.9\%$, $T=0.1\%$\\
cavity length $L=5\,km$\\
\hline
\end{tabular}
\caption{Basic parameters of the test setup}\label{tab:params}
\end{table}

\section{Field amplitudes for the aligned and resonant case}
In order to confirm that the test setup as described above has been
implemented correctly we record the amplitude or power of the light
fields in several locations, see table~\ref{tab:powers}.

\begin{table}[htb]
\centering
\begin{tabular}{|l r|l|l|l|}
\hline
  & & \Finesse & Analytic & OSCAR \\
\hline
sideband amplitude after EOM & [$\sqrt{\rm W}$]  &  $5\cdot 10^{-4 }$ & $5\cdot 10^{-4 }$ & $5\cdot 10^{-4 }$ \\
intra-cavity power & [W] & 89.92 &  89.92 & 89.92 \\
intra-cavity carrier amplitude & [$\sqrt{\rm W}$] & 9.483 & 9.483 & 9.483 \\
 WFS1 carrier amp. & [$\sqrt{\rm W}$] & 0.4528& 0.4528 & 0.4528\\
 WFS1 upper amp. & [$\sqrt{\rm W}$] & 2.5$\,\cdot 10^{-4 }$ & 2.5$\,\cdot 10^{-4 }$ & 2.5$\,\cdot 10^{-4 }$ \\
 WFS1 lower amp  & [$\sqrt{\rm W}$] & 2.5\,$\cdot 10^{-4 }$ & 2.5\,$\cdot 10^{-4 }$ & 2.5\,$\cdot 10^{-4 }$\\
 WFS2 carrier amp & [$\sqrt{\rm W}$] & 0.4528 &  0.4528 & 0.4528\\
 WFS2 upper amp.& [$\sqrt{\rm W}$] & 2.5 \,$\cdot 10^{-4 }$ & 2.5 \,$\cdot 10^{-4 }$  &  2.5 \,$\cdot 10^{-4 }$ \\
 WFS2 lower amp.& [$\sqrt{\rm W}$] & 2.5 \,$\cdot 10^{-4 }$  & 2.5 \,$\cdot 10^{-4 }$  & 2.5 \,$\cdot 10^{-4 }$ \\

\hline
\end{tabular}
\caption{Light powers and field amplitudes in various locations of the
optical layout.}\label{tab:powers}
\end{table}

\newpage

\section{Longitudinal error signal for small mirror offset}
As a first test of sensing and control signals we investigate the 
behavior of a Pound-Drever-Hall sensing:
The ETM is moved off-resonance by $0.1$\,nm. We compute the error signal from
the photo diode located in front of ITM, demodulated at 9\,MHz, in the
I-quadrature (defined by maximum signal). \Finesse and OSCAR results
for demodulated signals are multiplied by 2 to compensate
the built-in `mixer gain' of 0.5.

\begin{table}[htb]
\centering
\begin{tabular}{|l r|l|l|l|}
\hline
  & &\Finesse & Analytic & OSCAR \\
\hline
Circulating power & [W] &  88.8178        &       &  88.8176 \\
LSC demodulation phase & [deg] & 0.7574 & -0.7574 & 90.7552  \\
LSC signal in I phase (maximised)& [W]  &  $1.0484\cdot 10^{-4 }$ & $1.0484 \cdot 10^{-4}$ & $1.0483 \cdot 10^{-4 }$  \\
LSC signal in Q phase & [W] & $1.4548 \cdot 10^{-12}$&  $1.3587\cdot 10^{-20 }$ &  $-9.598\cdot 10^{-12}$ \\
\hline
\end{tabular}
\caption{Length sensing and control signal (LSC) taken at WFS1 position.}\label{tab:lsc1}
\end{table}


\section{Tilt of optical fields for small mirror misalignment}
Before we start computing wavefront sensor (WFS) signals, we want to
make sure that the tilt of the carrier and the sideband fields
are as expected.
\subsection{Wavefront tilt}
Compute the tilt of the wavefront on both WFSs as follows:
\begin{equation}
\phi_{\rm tilt}=\phi_{\rm sb} - \phi_{\rm carier}
\end{equation}
with $\phi$ being the phase of the respective field 
as the function of position on the WFS. 
Compute the slope of this for both WFSs, for the upper and lower
sideband, using  a) a misalignment of ITM by 0.1\,nrad and b)
a misalignment of ETM by 0.1\,nrad. The tilt of the wavefront
at the mirrors itself should be given by
\begin{equation}
 2 k \frac{180\,\mathrm{deg}}{\pi}\,0.1\,\mathrm{nrad}=6.7669173 \cdot 10^{-2 }\,\mathrm{deg/m}
\end{equation}

\begin{figure}
\centering
\includegraphics[width=8cm]{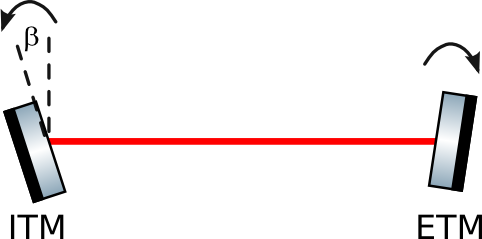}
\caption{Sign convention for misalignment angles. The illustration
  shows a misalignment of the ITM and ETM by a positive numerical
  value of the misalignment angle.}
\end{figure}

\Finesse results are for vertical misalignments (pitch) as discussed
at the workshop, to avoid sign flips upon reflection.

\begin{table}[htb]
\centering
a) ITM tilt \\
\begin{tabular}{|l r|r|r|r|}
\hline
  \multicolumn{1}{|c}{}&
  \multicolumn{1}{c}{}&
 \multicolumn{1}{|c}{\Finesse}&
\multicolumn{1}{|c}{Analytic}&
\multicolumn{1}{|c|}{OSCAR}\\
\hline
WFS1, upper sb - carrier& deg/m &$7.2752 \cdot 10^{-2 }$& $7.2752 \cdot 10^{-2 }$ $ \pm 0.062$ \% & $7.2778 \cdot 10^{-2 }$ \\
WFS1, lower sb - carrier&deg/m &$7.1301 \cdot 10^{-2 }$& $7.1301 \cdot 10^{-2 }$ $ \pm 0.067$ \% & $7.1339 \cdot 10^{-2 }$ \\
WFS2, upper sb - carrier&deg/m &$-1.5698 \cdot 10^{-3 }$& $-1.5698 \cdot 10^{-3 }$ $ \pm 0.110$ \% & $-1.5572 \cdot 10^{-3 }$  \\
WFS2, lower sb - carrier&deg/m &$-1.2906 \cdot 10^{-4 }$&  $-1.2909 \cdot 10^{-4 }$ $ \pm 3.600$ \% & $-1.2981 \cdot 10^{-4 }$ \\
\hline
\end{tabular}

\vspace{5mm}b) ETM tilt\\
\begin{tabular}{|l r|r|r|r|}
\hline
  \multicolumn{1}{|c}{}&
  \multicolumn{1}{c}{}&
 \multicolumn{1}{|c}{\Finesse}&
\multicolumn{1}{|c}{Analytic}&
\multicolumn{1}{|c|}{OSCAR}\\
\hline
WFS1, upper sb - carrier&deg/m &$-1.8428 \cdot 10^{-3 }$ & $-1.8429 \cdot 10^{-3 } \pm 0.120$ \% & $-1.8769 \cdot 10^{-3 }$\\
WFS1, lower sb - carrier&deg/m &$-1.5149 \cdot 10^{-4 }$ & $-1.5154 \cdot 10^{-4 } \pm 4.300$ \% &  $-2.0131 \cdot 10^{-4 }$ \\
WFS2, upper sb - carrier&deg/m &$-6.9613 \cdot 10^{-2 }$ & $-6.9613 \cdot 10^{-2 } \pm 0.061$ \% &  $-6.9623 \cdot 10^{-2 }$ \\
WFS2, lower sb - carrier&deg/m &$-7.1043 \cdot 10^{-2 }$ &  $-7.1043 \cdot 10^{-2 } \pm 0.056$ \% & $-7.1046 \cdot 10^{-2 }$  \\
\hline
\end{tabular}

\caption{Wavefront tilt for the beams for an ITM or ETM misalignment of 0.1\,
  nrad. Measurements taken at the two wavefront sensor
  positions. The \Finesse values have been computed as an average over
  one beam diameter. The analytic values are means along the misaligned (yaw) axis and
  the errors quoted represent the largest or smallest slope across the beam - out to $20~cm$ each side (this is the size of the virtual photodiode). Errors on wavefront tilt at the center of the beam are much smaller.}\label{tab:tilt}
\end{table}

\subsection{Beam propagation tilt}
\label{sec:beamtilt}
We can also estimate the tilt of the optical fields by comparing the
beam centers at two locations on the optical axis. For this we compute the beam center
on the WFSs and at temporary detectors, located (without any optical 
components in the path) 1\,km behind the respective WFS.
The beam center is estimated computing the `center of mass' of the beam
intensity on the detectors. The results are shows in table~\ref{tab:tilt2}
\begin{table}[H]
\centering
\begin{tabular}{|l|l r|r|r|r|}
\hline
 &  & &\Finesse  & OSCAR \\
\hline
\multirow{2}{*}{ITM tilt}  & WFS1 & [nrad] &  -2.390             &  -2.356 \\
 & WFS2 & [nrad] & 2.423  &     2.363  \\ 
 \hline
\multirow{2}{*}{ETM tilt}  & WFS1 & [nrad] &      2.833          &  2.790 \\
 & WFS2 & [nrad] &-2.848   &     -2.783  \\
\hline
\end{tabular}
\caption{Beam tilt of the carrier measured at WFS1 and WSF2 (based on the
  beam shift over 1\,km propagation).}\label{tab:tilt2}
\end{table}

\clearpage

\section{Wavefront sensor signal for mirror misalignment}
Now we compute the I and Q quadrature of WFS1 and WFS2 for
the same misalignments as above. The optimised demodulation phases are
shown in table~\ref{tab:ASC_demod}.

\begin{table}[H]
\centering
\begin{tabular}{|l|l|l|l|}
\hline
 & \Finesse & Analytic & OSCAR\\
\hline
WFS1, demodulation phase [deg] & 0.9859297 & -0.985930287636 & 90.98173\\
WFS2, demodulation phase [deg] & -36.3705156 & 36.370514867850581 & -126.4576 \\
\hline
\end{tabular}
\caption{Demodulation phase for the WFSs. They have been set as to
  maximise the ITM signal on WFS1 and the ETM signal on WFS2.}\label{tab:ASC_demod}
\end{table}%

\noindent Setting these phases we can the compute an alignment sensing matrix,
the \Finesse results are:
\begin{equation}
\begin{pmatrix}
\mathrm{WFS1}\\ \mathrm{WFS2}
\end{pmatrix}
=
\begin{pmatrix}
144.71 & -2.0033 \\
-1.849 &  -153.09\\
\end{pmatrix}
\cdot
\begin{pmatrix}
\mathrm{ITM}\\ \mathrm{ETM}
\end{pmatrix}
\end{equation}

\noindent A preliminary, analytically derived sensing matrix was
computed as:
\begin{equation}
\begin{pmatrix}
\mathrm{WFS1}\\ \mathrm{WFS2}
\end{pmatrix}
=
\begin{pmatrix}
144.4123 & -1.9992\\
-1.8457 & -152.823\\
\end{pmatrix}
\cdot
\begin{pmatrix}
\mathrm{ITM}\\ \mathrm{ETM}
\end{pmatrix}
\end{equation}
The discrepancy was shown to be due to numerical integration
limitations in MATLAB.  A grid with 5x higher resolution in both
dimensions resulted in the following result:
\begin{equation}
\begin{pmatrix}
\mathrm{WFS1}\\ \mathrm{WFS2}
\end{pmatrix}
=
\begin{pmatrix}
144.6978  &  -2.0032 \\
   -1.8489 & -153.0803 \\
\end{pmatrix}
\cdot
\begin{pmatrix}
\mathrm{ITM}\\ \mathrm{ETM}
\end{pmatrix}
\end{equation}
which is much closer to the \Finesse result. 

\vspace{5mm}
\noindent  The sensing matrix computed with OSCAR yields:
\begin{equation}
\begin{pmatrix}
\mathrm{WFS1}\\ \mathrm{WFS2}
\end{pmatrix}
=
\begin{pmatrix}
144.73  &  -1.9703 \\
   -1.8905 & -153.10 \\
\end{pmatrix}
\cdot
\begin{pmatrix}
\mathrm{ITM}\\ \mathrm{ETM}
\end{pmatrix}
\end{equation}

\begin{figure}[H]
\centering
	\includegraphics[scale=0.75]{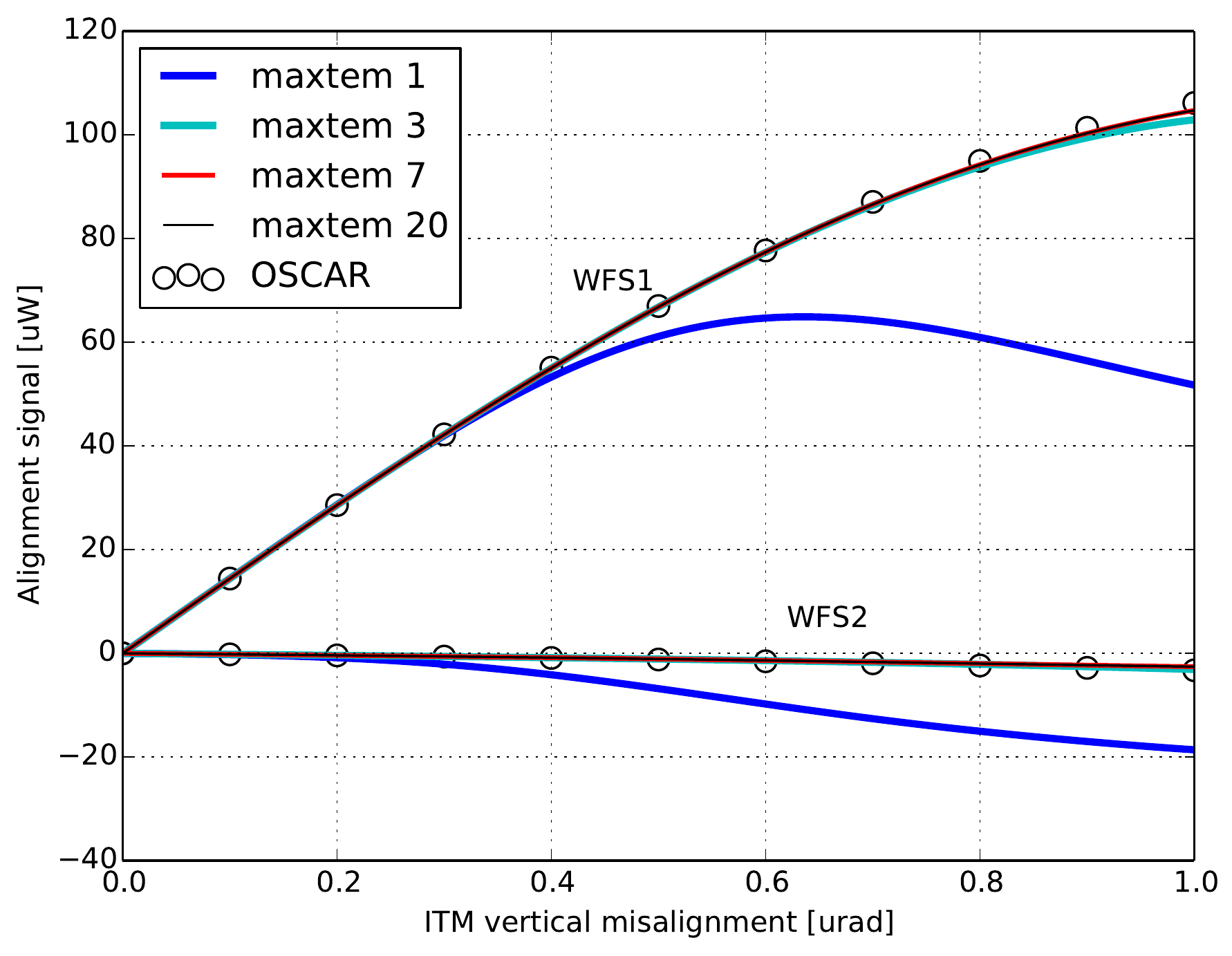}\\
	\includegraphics[scale=0.75]{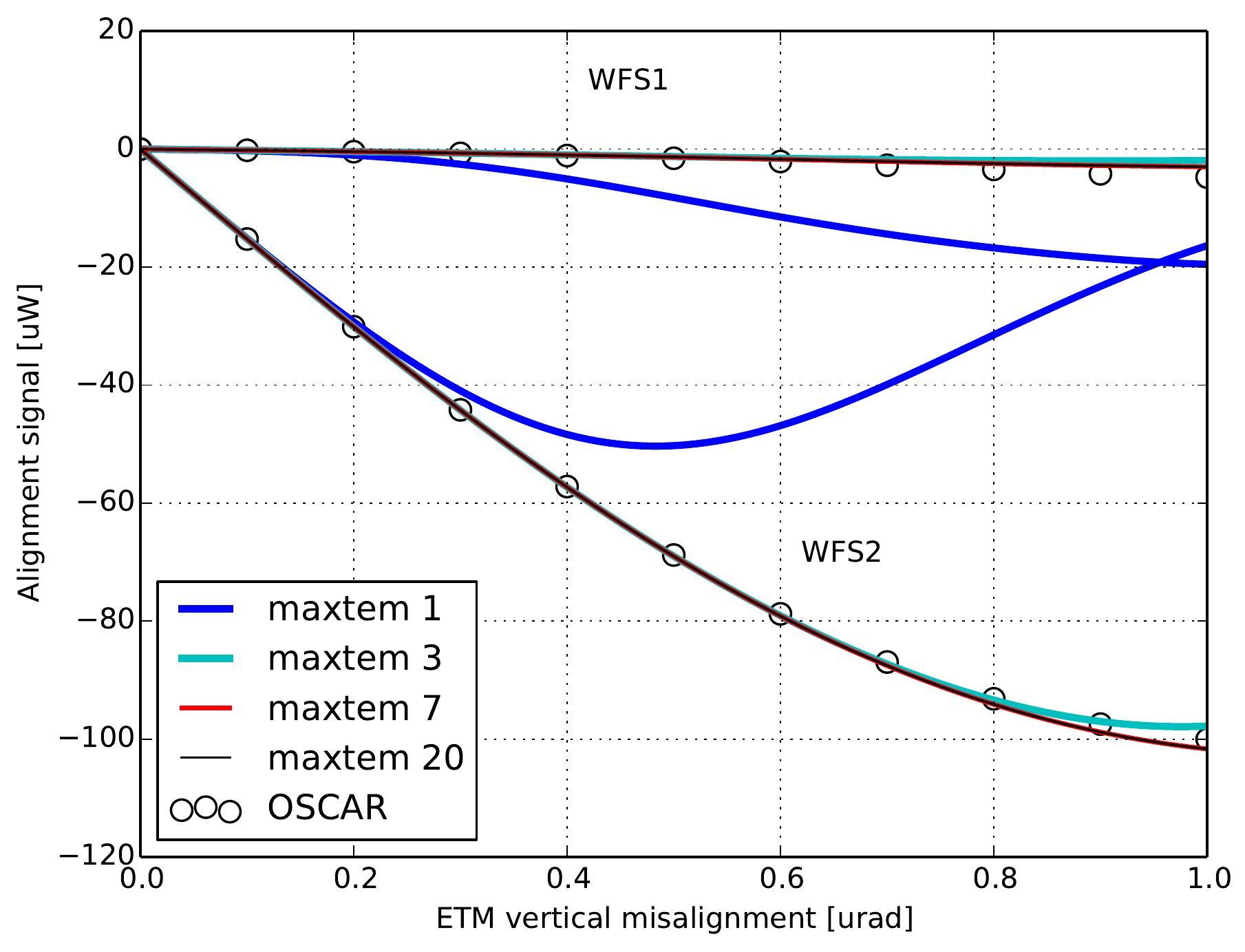}
\caption{Wavefront sensor signals for large misalignment of the input
and end mirror. The cavity is held on its resonance point longitudinally by
zeroing a Pound-Drever-Hall error signal. The \Finesse results are given for different settings of the
maximum higher order mode order, to show the convergence.}
\label{fig:large_misalign}
\end{figure}
\subsection{Large misalignments}
One of the more interesting tests is to model the wavefront sensor
(WFS) signals for larger misalignments. In this regime especially the
modal model is used outside the most simple approximation so that this
represents a much more generic and more meaningful test. This test
could not easily be performed with the analytic code. The comparison
between \Finesse \footnote{Please note that these
  \Finesse results require the new `split detector' definition is the
  {\tt kat.ini} as described in Appendix~\ref{sec:katini}.}
 and OSCAR is shown in figure~\ref{fig:large_misalign}.
The modal model converges quickly and the results from OSCAR
and \Finesse agree very well for misalignments below 0.4\,nrad,
and show a small but systematic difference for higher misalignments.

\section{Thermal lens}
The next step is to compute the alignment signals in the presence of 
a thermal lens in the ITM substrate. We assume the input mode to be 
the same as before the ITM thermal lens is introduced, therefore we 
must consider a mode mismatch between the input mode and the 
cavity eigenmode. 

In order to verify the optical setup we assume a 100\%
reflective ITM and compare the beam sizes of the wavefront sensors:

\begin{table}[th]
\centering
\begin{tabular}{|l|l|l|l|}
\hline
 & lens & \Finesse  & OSCAR\\
\hline
WFS1 & 50\,km&6.37232\,cm& 6.37364\,cm\\
WFS1 &  5\,km &6.37232\,cm&6.37499\,cm\\
WFS2 & 50\,km&8.09539\,cm & 8.09608\,cm\\
WFS2 &  5\,km& 19.3008\,cm&19.3031\,cm \\
\hline
\end{tabular}
\caption{Beam sizes on the wavefront senors for a thermal lens in the
  ITM (and assuming a 100\% reflective ITM).}\label{tab:lens_beamsize}
\end{table}%

\subsection{Beam propagation tilt with thermal lens}
Similar to section~\ref{sec:beamtilt} we
estimate the tilt of the optical fields using a 
`center of mass' computation. The results in table~\ref{tab:tilt3}
show differences between OSCAR and \Finesse. The reasons
for that are not clear to us. Consistency checks were done
with \Finesse and OSCAR and did not reveal any information
on which code produces more accurate results.
\begin{table}[H]
\centering
\begin{tabular}{|l|l|l r|r|r|}
\hline
 &  lens [km]& & &\Finesse  & OSCAR \\
\hline
\multirow{2}{*}{ITM tilt}  & 50 & WFS1 & [nrad] &-2.366        &  -1.763\\
 & 50 & WFS2 & [nrad] & 2.309  &   2.027  \\ 
 \hline
\multirow{2}{*}{ETM tilt}  & 50 & WFS1 & [nrad] &    2.883        &  2.262 \\
& 50 & WFS2 & [nrad] & -2.801   &   -2.823  \\
\hline
\multirow{2}{*}{ITM tilt}  & 5 & WFS1 & [nrad] &-0.843       &  -0.625\\
 & 5 & WFS2 & [nrad] & 2.043  &   0.737  \\ 
 \hline
\multirow{2}{*}{ETM tilt}  & 5 & WFS1 & [nrad] &   0.952        &  0.509 \\
& 5 & WFS2 & [nrad] & -2.893  &   -1.32  \\
\hline
\end{tabular}
\caption{Beam tilt of the carrier measured at WFS1 and WSF2, now with
  thermal lens in the ITM (\Finesse results have been obtained with
  \texttt{maxtem 8} for the 50\,km lens and \texttt{maxtem 29} for
  the 5\,km lens).}\label{tab:tilt3}
\end{table}

\begin{figure}[H]
\centering
	\includegraphics[width=0.75\textwidth]{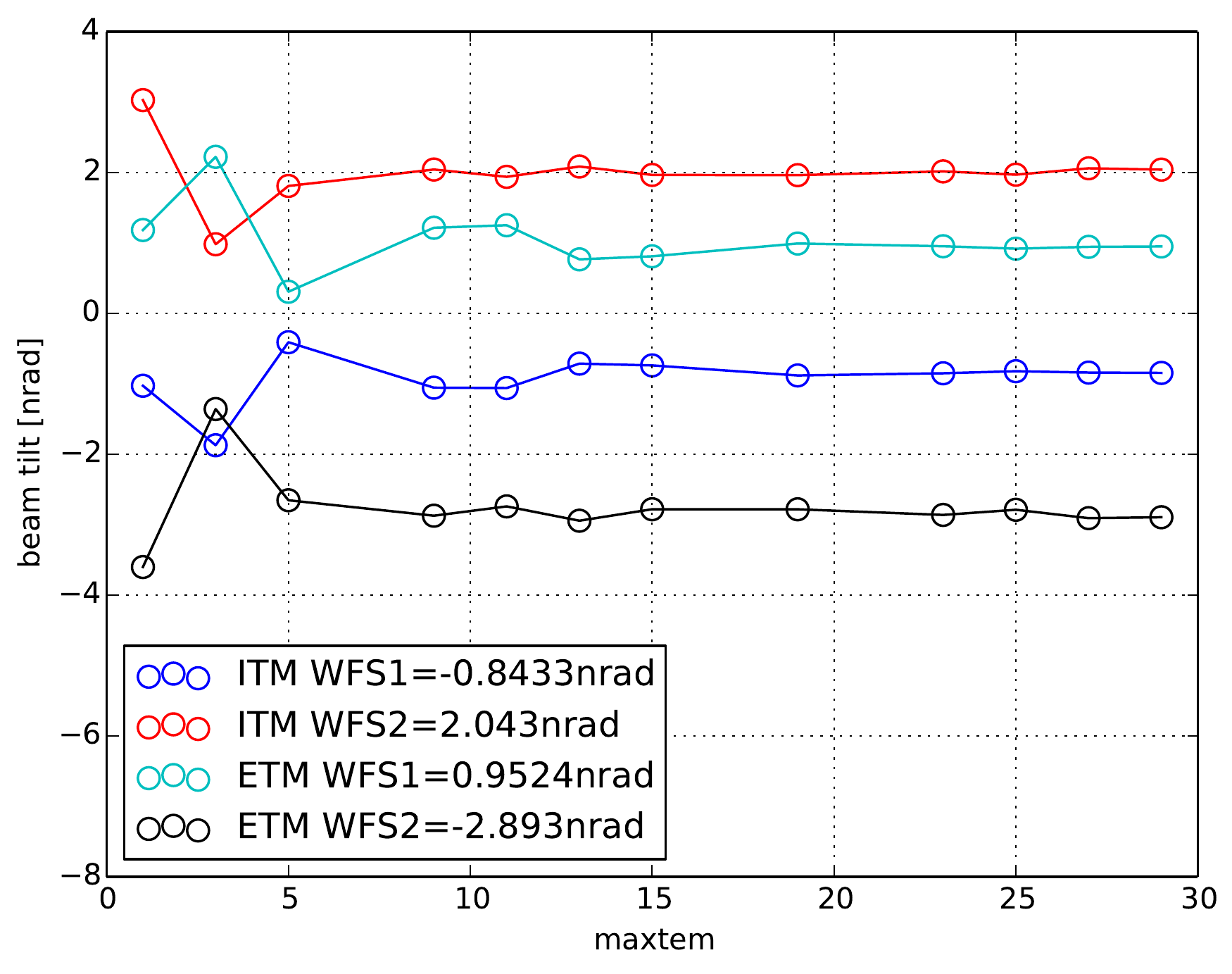}\\
\caption{Beam propagation tilt for a 5\,km thermal lens a function of
 the  'maxtem' setting in \Finesse. Convergence is clearly noticeable,
 however a `wobble' in the results remains,
 possibly due to increasing numerical errors.}
\label{fig:gravity_lens}
\end{figure}

\subsection{Manually setting the beam parameter in \Finesse}
The input mode was previously automatically matched to the 
cavity eigenmode using the \texttt{cav} command in \Finesse, but 
in order the simulate the mode mismatch we must manually set the input mode parameters using 
the \texttt{gauss} command. 

In addition we might consider setting a beam parameter at the pick-off
beam splitter:
With the input mode and the cavity eigenmode
not matched, there is no beam parameter in which the light returning
from the cavity
can be described as a fundamental beam. Thus higher-order modes are
necessary to describe the reflected beam. Especially the thermal
lens of 5\,km poses a challenge as the beam parameters of the
input beam (assuming for a moment a fully reflective ITM) and that of the cavity eigenmode
(transmitted through the ITM) differ strongly. Measuring these beam parameters at the pick-off
beam splitter yields:
\begin{itemize}
\item beam parameter of reflected input beam: $w_0=6.37$\,cm, $z<1$\,mm
\item beam parameter of cavity eigenmode: $w_0=1.3$\,cm, $z=2.395$\,km
\end{itemize}
From the section `Limits to the paraxial approximation' in the
\Finesse manual we expect this system to be outside the range in 
which simple paraxial models can be used. By manually setting a beam
parameter  ($w_0=2.8$\,cm, $z=1.677$\,km) at the pick-off beam
splitter as a compromise between the parameters measured above we
can reduce the mode-mismatch in the calculation. This method has been 
used to compute the data shown in figure~\ref{fig:gravity_lens} and in the following.

\subsection{Wavefront sensor signal with thermal lens}
\begin{figure}[H]
\centering
	\includegraphics[width=0.7\textwidth]{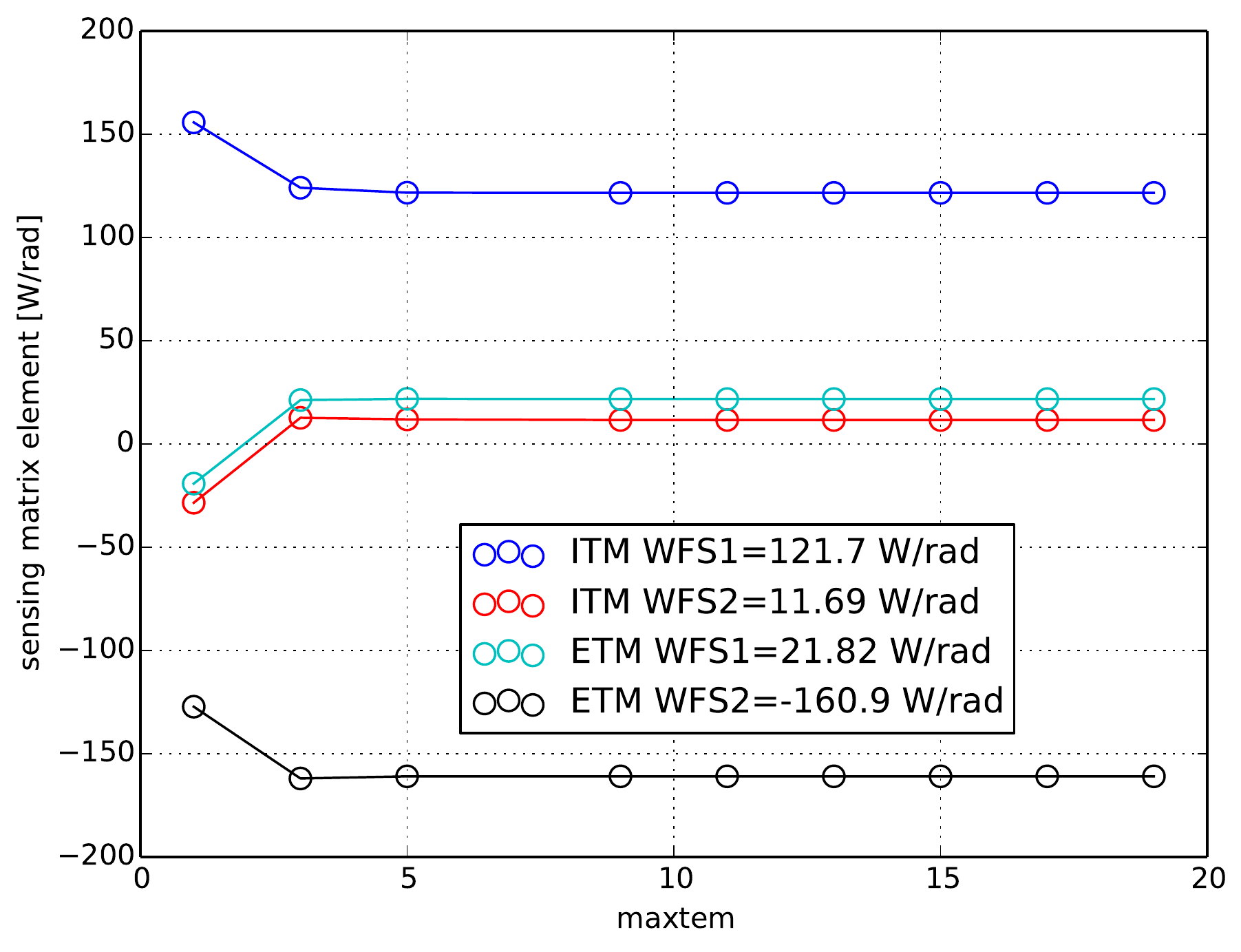}\\
\caption{Convergence of \Finesse alignment matrix calculation in the
  presence of a thermal lens of $f=50$\,km.}
\label{fig:finesse_lens1}
\end{figure}

\begin{figure}[ht]
\centering
	\includegraphics[width=0.7\textwidth]{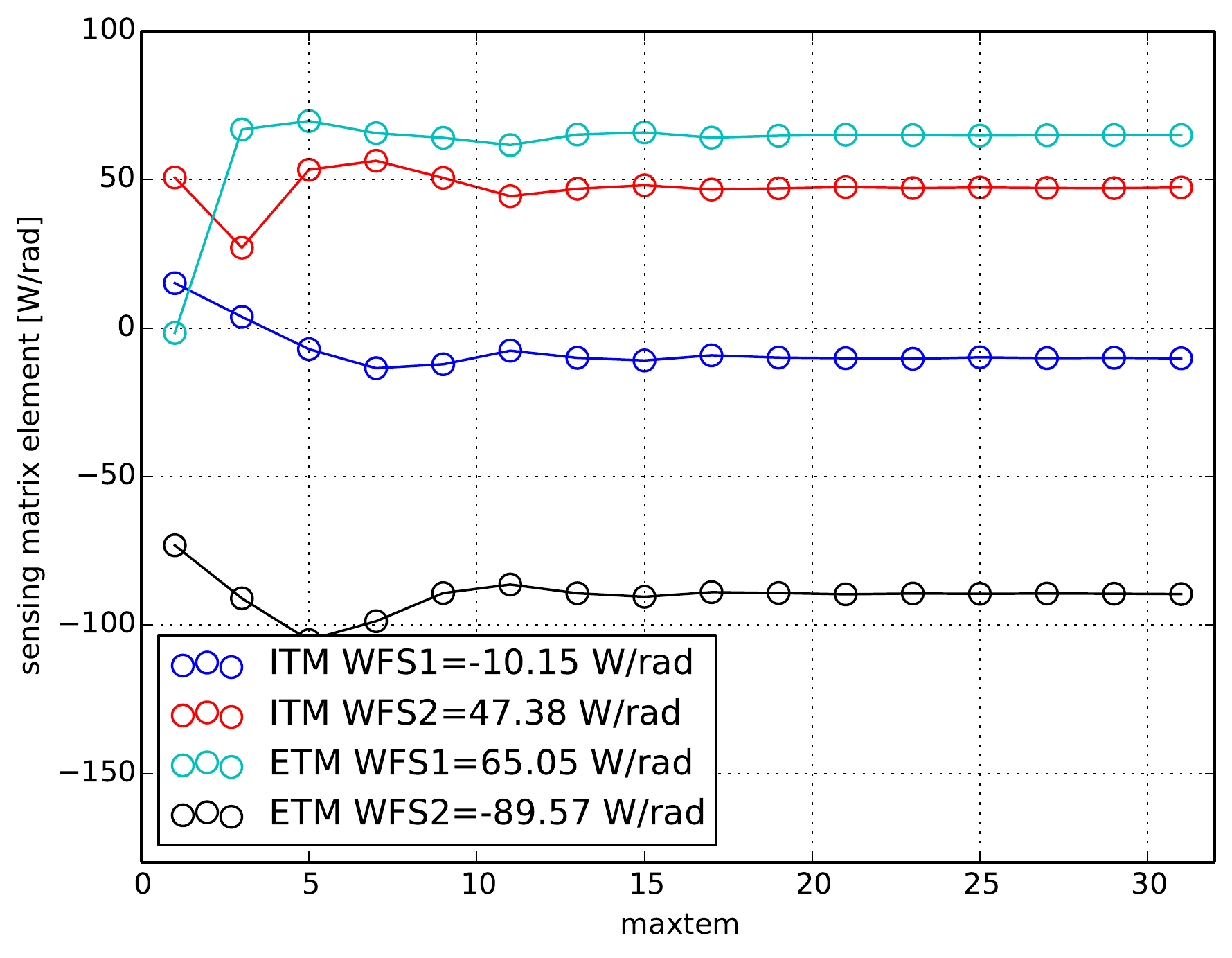}\\
\caption{Convergence of \Finesse alignment matrix calculation in the
  presence of a thermal lens of $f=5$\,km.}
\label{fig:finesse_lens2}
\end{figure}

Next we calculate the sensing  matrix for small misalignments as before,
but now in the presence of the thermal lens.  The mode mismatch causes coupling into a wider range 
of higher-order modes than before, so it is necessary to use a high \texttt{maxtem} value. In 
order to check this, we calculate the sensing matrix for a range of \texttt{maxtem} values and look for 
convergence (see figures \ref{fig:finesse_lens1} and ~\ref{fig:finesse_lens2}). For the 50\,km lens
case, convergence is reached by \texttt{maxtem} 7 or so, the 5\,km
lens requires at least \texttt{maxtem} 15.

\subsubsection{50\,km lens in the ITM substrate}

\noindent Sensing matrix computed with \Finesse:
\begin{equation}
\begin{pmatrix}
\mathrm{WFS1}\\ \mathrm{WFS2}
\end{pmatrix}
=
\begin{pmatrix}
121.7 & 21.82 \\
11.69 &  -160.9\\
\end{pmatrix}
\cdot
\begin{pmatrix}
\mathrm{ITM}\\ \mathrm{ETM}
\end{pmatrix}
\end{equation}

\noindent Results obtained with OSCAR:
\begin{equation}
\begin{pmatrix}
\mathrm{WFS1}\\ \mathrm{WFS2}
\end{pmatrix}
=
\begin{pmatrix}
121.98  &  21.50 \\
  11.51 & -160.55 \\
\end{pmatrix}
\cdot
\begin{pmatrix}
\mathrm{ITM}\\ \mathrm{ETM}
\end{pmatrix}
\end{equation}

\noindent Results obtained with the analytic code:
\begin{equation}
\begin{pmatrix}
\mathrm{WFS1}\\ \mathrm{WFS2}
\end{pmatrix}
=
\begin{pmatrix}
121.71  &  21.74 \\
  11.65 & -160.78 \\
\end{pmatrix}
\cdot
\begin{pmatrix}
\mathrm{ITM}\\ \mathrm{ETM}
\end{pmatrix}
\end{equation}

\subsubsection{5\,km lens in the ITM substrate}
Now we compute the alignment sensing matrix with a 5\,km thermal lens in the ITM substrate. 
This is approximately the maximum focal power that can be expected in the 
uncompensated aLIGO ITM.

\vspace{5mm}
\noindent Sensing matrix computed with \Finesse at \texttt{maxtem} 31:
\begin{equation}
\begin{pmatrix}
\mathrm{WFS1}\\ \mathrm{WFS2}
\end{pmatrix}
=
\begin{pmatrix}
-10.15 & 65.05\\
47.38 &  -89.57\\
\end{pmatrix}
\cdot
\begin{pmatrix}
\mathrm{ITM}\\ \mathrm{ETM}
\end{pmatrix}
\end{equation}

\noindent Results obtained with OSCAR:
\begin{equation}
\begin{pmatrix}
\mathrm{WFS1}\\ \mathrm{WFS2}
\end{pmatrix}
=
\begin{pmatrix}
-0.309  &  54.96 \\
  38.46 & -126.62 \\
\end{pmatrix}
\cdot
\begin{pmatrix}
\mathrm{ITM}\\ \mathrm{ETM}
\end{pmatrix}
\end{equation}

\noindent Results obtained with the analytic code:
\begin{equation}
\begin{pmatrix}
\mathrm{WFS1}\\ \mathrm{WFS2}
\end{pmatrix}
=
\begin{pmatrix}
-13.43  &  68.35 \\
  50.94 & -94.31 \\
\end{pmatrix}
\cdot
\begin{pmatrix}
\mathrm{ITM}\\ \mathrm{ETM}
\end{pmatrix}
\end{equation}

\subsubsection{Conclusion}
The results summarized in table~\ref{tab:asc_full} 
show some significant differences
between the different methods, especially the  OSCAR 
results include matrix elements which are different by
more than a factor of 3.
This is not very surprising though because
already the results for the beam propagation tilt with a 5\,km
thermal lens showed similarly large  differences.
Unfortunately we do not have a reference result
and thus do not know if one of the three results is correct.

\begin{table}[H]
\centering
\begin{tabular}{|l|r|r|r|r|r|r|r|r|}
\hline
&\multicolumn{4}{|c}{50\,km lens} &\multicolumn{4}{|c|}{5\,km lens}\\
\hline
 & \multicolumn{2}{c}{WFS1} &\multicolumn{2}{|c|}{WFS2} 
 & \multicolumn{2}{c}{WFS1} &\multicolumn{2}{|c|}{WFS2} \\
\cline{2-9}
 &  ITM & ETM & ITM &ETM
 &  ITM & ETM & ITM &ETM\\
\cline{2-9}
\Finesse & 121.70 & 21.82 & 11.69 & -160.90 
 & -10.15 & 65.05 & 47.38 & -89.57 \\
OSCAR & 121.98 & 21.50 & 11.51 & -160.55
 & -0.31 & 54.96 & 38.46 & -126.62 \\ 
Analytic & 121.71 & 21.74 & 11.65& -160.78
 &-13.43 & 68.35& 50.94 & -94.31\\
\hline
\end{tabular}
\caption{Summary table, showing all sensing matrix elements as
  computed will three different methods for the 50\,km lens and for
  the 5\,km lens.}\label{tab:asc_full}
\end{table}


\clearpage
\begin{appendices}
\section{\Finesse files}
All the \Finesse results presented in this document
have been computed starting with the base file
shown below. For each simulation the
extra detectors and commands
have been added using a set of script files. This approach has the 
advantage of creating a track record of the entire process: 
all the results shown in this document can be 
reproduced at any given time, simply by re-running the scripts.
The script files contain (and thus document) all changes to the
\Finesse input file as well as any post-processing of the
results.

Originally the scripts had been written in Matlab, using the
Simtools package~\cite{simtools}. Recently we have developed
the Python package PyKat~\cite{pykat} which provides a more
powerful and effective way to write script files for running
\Finesse simulations. A new set of PyKat scripts provided all
the results shown in this document has been created and
is available as an example in the PyKat
package itself. PyKat can be downloaded at \url{http://www.gwoptics.org/pykat}.

\vspace{5mm}
\noindent The base file containing the main optical setup:
\begin{finesse}
l psl 1.0 0 npsl 
mod EOM 9M 0.001 1 pm 0 npsl nEOM1
s s1 0 nEOM1 npo1

bs1 po 0.5 0 0 45 npo1 dump npo2 nWFS1 % 50:50 pick-off mirror
s s2 0 npo2 nL1

lens ITM_TL 10000G nL1 nL2  % thermal lens in ITM
s ITMsub 0 nL2 nITM1  

m1 ITM 0.02 0 0 nITM1 nITM2  
attr ITM Rc -2500  
s s_cav 5000 nITM2 nETM1    
m1 ETM 0.001 0 0 nETM1 nETM2
attr ETM Rc 2700   
cav c1 ITM nITM2 ETM nETM1

s spo1 1n nWFS1 nL1_in
lens L1 1250 nL1_in nL1_out 
s spo2 5000 nL1_out nWFS2 

phase 0
\end{finesse}

\newpage
\section{Split detectors definition in the {\tt kat.ini} file of \Finesse}\label{sec:katini}
During the simulation workshop we realised that the definition of split
detectors (WFS) in the default {\tt kat.ini} file distributed with
\Finesse until version 1.0 was not correct. The original definition 
looks as follows:
\begin{finesse}
PDTYPE y-split
x 0 x 1  1.0
x 0 x 3  1.0
x 0 x 5  1.0
x 0 x 7  1.0
x 0 x 9  1.0
x 0 x 11 1.0
x 0 x 13 1.0
x 0 x 15 1.0
x 2 x 1  1.0
x 2 x 3  1.0
x 2 x 5  1.0
x 2 x 7  1.0
x 2 x 9  1.0
x 2 x 11 1.0
x 2 x 13 1.0
x 2 x 15 1.0
x 4 x 1  1.0
x 4 x 3  1.0
x 4 x 5  1.0
x 4 x 7  1.0
x 4 x 9  1.0
...
\end{finesse}

In order to continue this document a correct sequence of beat
coefficients has been derived and is available now. The derivation is
described in the new \Finesse manual and the resulting coefficients
list begins like this:
\begin{finesse}
PDTYPE y-split
x  0 x  1 0.797884560802865
x  0 x  3 -0.32573500793528
x  0 x  5 0.218509686118416
x  0 x  7 -0.168583882836184
x  0 x  9 0.139074607877595
x  0 x 11 -0.119342192152758
x  0 x 13 0.105105246952603
x  0 x 15 -0.0942883643366146
x  2 x  1 0.564189583547756
x  2 x  3 0.690988298942671
x  2 x  5 -0.257516134682126
x  2 x  7 0.166889529453113
x  2 x  9 -0.126437912127138
x  2 x 11 0.103140489653528
x  2 x 13 -0.0878334751963773
x  2 x 15 0.0769291636262473
x  4 x  1 -0.16286750396764
x  4 x  3 0.598413420602149
x  4 x  5 0.669046543557288
x  4 x  7 -0.240884286886713
x  4 x  9 0.153297821464995
...
\end{finesse}

\end{appendices}

\end{document}